\documentclass[a4paper,10pt]{article}
\usepackage{graphicx}
\usepackage{dcolumn}
\usepackage{bm}
\usepackage{amsmath}

\title{One-step replica symmetry breaking solution for fermionic Ising spin glass in a transverse field}

\author{F. M. Zimmer, S.G. Magalhaes\footnote{ggarcia@ccne.ufsm.br}%
\\
\\
{\it Lab. de Mec\^anica Estat\'\i stica e Teoria da Mat\'eria Condensada,}
\\
{\it Dep. F\'\i sica, UFSM,
97105-900 Santa Maria, RS, Brazil}}
\date{}
\begin{document}
\maketitle

\begin{abstract}
The fermionic Ising spin glass models in a transverse field are investigated in a
Grassmann path integral formalism.
The Parisi's scheme of one-step replica symmetry
breaking (RSB)
is used within the static ansatz. 
 This formalism has already been applied in a theory in which $m$ (Parisi's block-size parameter)
is taken as a constant \cite{eduardo}. Now, it is extended to consider $m$ as a variational parameter. In this case,
the results show that  RSB is present when $T\rightarrow0$, in which the system is driven 
by quantum fluctuations.    
\end{abstract}

The infinite range Ising spin glass (ISG) in a transverse field has been extensively studied as a theoretical quantum spin glass model. 
The strength of quantum fluctuations is adjusted by the transverse magnetic field. 
An experimental 
realization of this model is the quantum spin glass system $LiHo_xY_{1-x}F_4$ \cite{experimental}. 
Moreover, there are now several disordered Cerium and Uranium alloys \cite{experimental2} 
in which it is possible to find a spin glass phase (SG) as well as a Kondo behavior 
around a quantum critical point (QCP). 
It is not our intention to discuss this particular complicated problem. However,  a proper 
fermionic formulation of the SG problem in a transverse field can also be useful for this 
mentioned class of strongly correlated problems \cite{theoretical}. 

Several techniques have been used to treat the ISG in a transverse field. 
Nevertheless, they 
show 
some controversial results 
concerning the local stability of replica symmetry solution \cite{thirumalai,buttner,goldschmidt,buttner1,koreanos,Alba2002}. 
For instance, it is  an open question 
whether or not the quantum tunneling through the barriers 
between the 
many degenerated thermodynamic states 
in the free energy landscape 
is able to restore the replica symmetry (RS). 
The quantum ISG model treated by the Trotter-Suzuki formalism in the static approximation \cite{braymoore} has 
shown stable RS solutions in a small region close to freezing temperature $T_c$ \cite{thirumalai}.
However, the replica symmetry breaking (RSB) is found in whole SG phase 
when the Trotter-Suzuki formalism is used without the static approximation,  in 
which the dynamic spin self-interaction is treated by a numerical calculation 
\cite{buttner}.
For this case, the ordered SG phase has been investigated by one-step RSB solution, 
\cite{goldschmidt,buttner1} in which it is suggested that the quantum tunneling does not 
restore the RS.
On the other hand, this quantum problem  has also been investigated by using imaginary-time replica formalism under the 
static ansatz.  This approach has shown that the RS quantum SG phase is 
stable in almost 
whole SG phase \cite{koreanos}.
 Furthermore, the SG quantum rotors as well as the transverse field ISG have been
studied within a Landau theory in which the dynamic effects are considered in an analytical approximation near the critical temperature \cite{sachdev}. In this treatment, the RSB occurs in the SG phase at finite temperatures and it 
suggests suppression of the RSB 
when $T\rightarrow 0$.

Recently, the fermionic ISG in a transverse field $\Gamma$ has been studied using 
functional integral formalism with Grassmann variables \cite{Alba2002}.
This  problem has been treated within the static approximation. 
The phase diagram
found  in 
Ref. \cite{Alba2002} shows that the $T_c$ decreases toward a QCP
when $\Gamma$ 
reaches a critical value. 
In this formalism, the RS solution is unstable in whole SG phase \cite{Alba2002}  in disagreement with 
Ref. \cite{koreanos}. 
In a subsequent work still using the static approximation, 
the replica symmetry  has been broken within 
the one-step RSB 
like-Parisi's scheme 
\cite{parisi}.
Nonetheless, 
the block-size parameter $m$ has been taken as a temperature independent \cite{eduardo}.
The results within this particular RSB treatment suggest that the RS is restored at $T=0$ when $\Gamma>0$ \cite{eduardo}.
However, it arises some questions   
whether
$m$  
is taken as a variational parameter  
as in
the original Parisi's scheme \cite{parisi} for $\Gamma>0$. 
Are the results using the original Parisi's scheme of RSB  
changed significantly when compared 
with those obtained in 
Ref. \cite{eduardo} for $\Gamma>0$?
Are the quantum effects able to restore the RS in this fermionic formulation?

The purpose of the present work  
is to investigate the one-step RSB 
in the fermionic ISG in a transverse field using the original Parisi's scheme. 
In particular, we  
compare it with the alternative procedure
of the same problem 
proposed in Ref. \cite{eduardo}. 
The problem is formulated in a path integral formalism as in 
Ref. \cite{eduardo}.   The static approximation is used 
to treat the spin-spin correlation functions.
However, the parameter $m$ 
is taken as a saddle-point equation, which is the main 
difference between this work and 
Ref. \cite{eduardo}. 
In our quantum spin glass treatment, the static approximation is an important point.  
It is recognized that the static approximation 
can yield 
inaccurate quantitative results at very low temperature. 
However, because of a lack of a suitable analytical method to 
go beyond the static ansatz in whole SG phase in this fermionic formalism, 
there are still several quite recent papers studying the ISG in transverse 
field 
within this particular level of 
approximation \cite{Alba2002,eduardo}.
This paper reports results for the one-step RSB order parameters, 
free energy and entropy. 
Our  results agree 
partially 
with those of 
Ref. \cite{eduardo}. 
Nevertheless, the RS is not 
restored at $T\rightarrow0$, which is in agreement with Refs. \cite{goldschmidt,buttner1}.  

The infinite range ISG in a transverse magnetic field $\Gamma$ is described by 
the Hamiltonian:
\begin{equation}
\hat{ H}= -\sum_{i\neq j} J_{ij}\hat{S}_{i}^{z} \hat{S}_{j}^{z} 
-2 \Gamma \sum_{i} \hat{S}_{i}^{x}\label{ham}
\end{equation}
\noindent
where the sums are run over the $N$ sites of a lattice. The exchange interaction 
$J_{ij}$ among all pairs of spins is a random variable that follows
a Gaussian probability distribution with mean equal to zero and variance $16J^2$.
The spin operators in Eq. (\ref{ham}) are defined in terms of fermion operators
\cite{Alba2002}:
$\hat S_{i}^{z}=\frac{1}{2}[\hat{n}_{i\uparrow}-\hat{n}_{i\downarrow}],~
\hat S_{i}^{x}=\frac{1}{2}[c_{i\uparrow}^{\dagger}c_{i\downarrow}
+c_{i\downarrow}^{\dagger}c_{i\uparrow}]$
where $\hat{n}_{i\sigma}=c_{i\sigma}^{\dagger}c_{i\sigma}$ gives the number of fermions 
at site $i$ with spin projection $\sigma=\uparrow$ or $\downarrow$. $c_{i\sigma}^{\dagger}$ and $c_{i\sigma}$ are respectively the
fermion creation and annihilation operators.

The Hamiltonian (\ref{ham}) is defined on a fermionic space where the operator $S^{z}$ has two magnetic
eigenvalues
and 
two non-magnetic eigenvalues (empty and double occupied site). 
Therefore, 
there are two states that do not belong
to the usual spin space. 
However, it is possible to express the partition function of a 
spin system (2S-model) in terms of the partition function of a corresponding fermionic system 
by using a restriction that considers only sites
occupied by one fermion \cite{wener}.
This work considers two models: the 4S-model without any restriction 
about 
the non-magnetic states 
and the 2S-model in which the non-magnetic states are forbidden. 
The restriction in the 2S-model is obtained by using the Kronecker delta function \cite{wener}. 
Therefore, the partition functions for both models are written in a compact form as \cite{Alba2002}:
\begin{equation}
Z\{y\}=\int D(\phi^* \phi)\prod_{j}^{}\frac{1}{2\pi}\int_{0}^{2\pi}dx_j\mbox{e}^{-y_j}\mbox{e}^{A(y)}
\end{equation}
with the action
\begin{equation}
\begin{split}
A\{y\}=\int_{0}^{\beta}d\tau
\{\sum_{j,\sigma}\phi_{j\sigma}^{*}(\tau)
 [\frac{\partial}{\partial \tau} +y_j]
 \phi_{j\sigma}(\tau)
 \\
  - H\left(\phi_{j\sigma}^{*}
 (\tau),\phi_{j\sigma}(\tau)\right) \}
\end{split}
\label{action},
\end{equation}
$\beta=1/T$ ($T$ is the temperature), $y_{j}=ix_j$ for the $2S$-model or $y_{j}=0$ for 
the $4S$-model, which corresponds to the half-filling situation. 

In Eq. (\ref{action}), the Fourier decomposition of the time-dependent quantities is employed. 
The configurational averaged free energy per site is obtained by using the replica method: 
$\beta F
=-\frac{1}{N}\displaystyle\lim_{n\rightarrow 0}(\langle Z\{y\}^n \rangle_{J_{ij}} - 1)/n$
where the replicated partition function is   
\begin{equation}
\begin{split}
\langle Z\{y\}^n\rangle_{J_{ij}}=\int dU
\exp[-N(\frac{\beta^2J^2}{2}\sum_{\alpha,\gamma}q_{\alpha,\gamma}^{2}
\\
+\ln\Lambda\{y\})]
\end{split}
\end{equation}
with $\int dU=\int_{-\infty}^{\infty}\prod_{\alpha,\gamma}dq_{\alpha\gamma}(\beta J\sqrt{N/2\pi})$
where $\alpha$ and $\gamma$ are replica indices running from 1 to $n$. It is
assumed the static approximation, therefore:
%
\begin{equation}
\Lambda\{y\}=\prod_{\alpha}\int_{0}^{2\pi}\frac{dx_{\alpha}}{2\pi}\mbox{e}^{-y_{\alpha}}
\int D[\phi_{\alpha}^{*},\phi_{\alpha}]\exp[\bar{H}],
\label{efetivo}
\end{equation}
\begin{equation}
\bar{H}=\sum_{\alpha} A_{0\Gamma}^{\alpha} + 4\beta^{2}J^{2}\sum_{\alpha,\gamma}
q_{\alpha\gamma}S_{\alpha}^{z}S_{\gamma}^{z}  
\label{heff}
\end{equation}
with the definitions:
\begin{eqnarray}
A_{0\Gamma}^{\alpha}=\sum_{\omega}\underline{\varphi}_{\alpha}^{\dagger}(\omega)(i\omega+y_\alpha+
\beta\Gamma\underline{\sigma}^{x})
\underline{\varphi}_{\alpha}(\omega),
\\
S_{\alpha}^{z}=\frac{1}{2}\sum_{\omega}
\underline{\varphi}_\alpha(\omega)\underline{\sigma}^{z}\underline{\varphi}_\alpha(\omega),
\end{eqnarray}
$\underline{\varphi}_{\alpha}^{\dagger}(\omega)=\left(\phi_{\uparrow\alpha}^{*}(\omega)~~~
\phi_{\downarrow\alpha}^{*}(\omega)\right)$ is a Grassmann spinor, $\underline{\sigma}^x$ 
and $\underline{\sigma}^z$ are the Pauli matrices, and $\omega$ denotes fermionic Matsubara 
frequencies.

The set of integrals in $\int dU$ has been exactly performed in the thermodynamic 
limit by the steepest descent method.
The simplest conjecture to the replica matrix $\{Q\}$ for overlaps in the spin space 
is the replica symmetry ansatz (RS). However, it produces an unstable solution in 
the ordered spin glass phase \cite{eduardo}.  An attempt to solve this problem 
is the so called method of replica symmetry breaking (RSB) \cite{parisi}.

In the Parisi's scheme of one-step RSB (1S-RSB),
the replica matrix $\{Q\}$ is parameterized
dividing the $n$ replicas into $n/m$ groups with $m$ replicas in each one, such as:
$q_{\alpha\alpha}=r$ and
\begin{equation}
\begin{split}
q_{\alpha\gamma}&=q_{1} \mbox{ if } I(\alpha/m)=I(\gamma/m)
\\
q_{\alpha\gamma}&=q_{0} \mbox{ if } I(\alpha/m)\neq I(\gamma/m)
\label{parametrization}
\end{split}
\end{equation}
where $I(x)$ gives the smallest integer which is greater than or equal to $x$. 
The parameter $r$ is the replica-diagonal spin-spin correlation $r=\langle S^{\alpha}
S^{\alpha}\rangle$. The physical meaning of $m$ is that two different spin glass order parameters
$q_0$ and $q_1$ are mixed up in the ratio $m:1-m$.

The parametrization (\ref{parametrization}) is used to sum over the replica index. It produces quadratic terms in 
Eq. (\ref{heff}) that are linearized introducing new auxiliary fields in Eq. (\ref{efetivo}).
The functional Grassmann integral is now performed and the sum over the Matsubara 
frequencies can be evaluated like references \cite{Alba2002,eduardo}. Finally, the restriction
condition over the number of states admitted in each model is explicitly adopted. 
This procedure results in the following expression to the free energy: 
\begin{equation}
\begin{split}
\beta F_s=\frac{\beta^2 J^2}{2}[m(q_1^2-q_0^2)+ r^2-q_1^2]\\
- \frac{1}{m}\int_{-\infty}^{\infty} Dz 
\ln\int_{-\infty}^{\infty} Dv (2 K_s)^m
\label{potencial}
\end{split}
\end{equation}
with 
\begin{equation}
K_{s}=\frac{s-2}{2}+ \int_{-\infty}^{\infty} D\xi\cosh(\beta\sqrt{\Delta})
\label{}
\end{equation}
where $s~(=2$ or $4)$ represents the number of states admitted in each model, $Dx=dx~\mbox{e}^{-x^2/2}/\sqrt{2\pi}$ ($x=z, v \mbox{ or } \xi$), 
$\Delta=h^2+\Gamma^2$, $h=J(\sqrt{2q_0}z + \sqrt{2(q_1-q_0)}v + \sqrt{2(r-q_1)}\xi)$ and the
parameters $q_0,~q_1,~r$ and $m$ are given by the extreme condition of the free energy (\ref{potencial}).

The set of saddle-point equations for the parameters $q_0,~q_1,~r$ and $m$  can  
numerically be solved. Here, the equation for $m$ is self-consistently solved
with the parameters $q_0,~q_1$ and $r$ \cite{parisi}
instead of 
Ref. \cite{eduardo} where $m$ is taken as a constant value belong to the interval $0< m <1$.

\begin{figure}[t!] 
\includegraphics[angle=270, width=8.5cm]{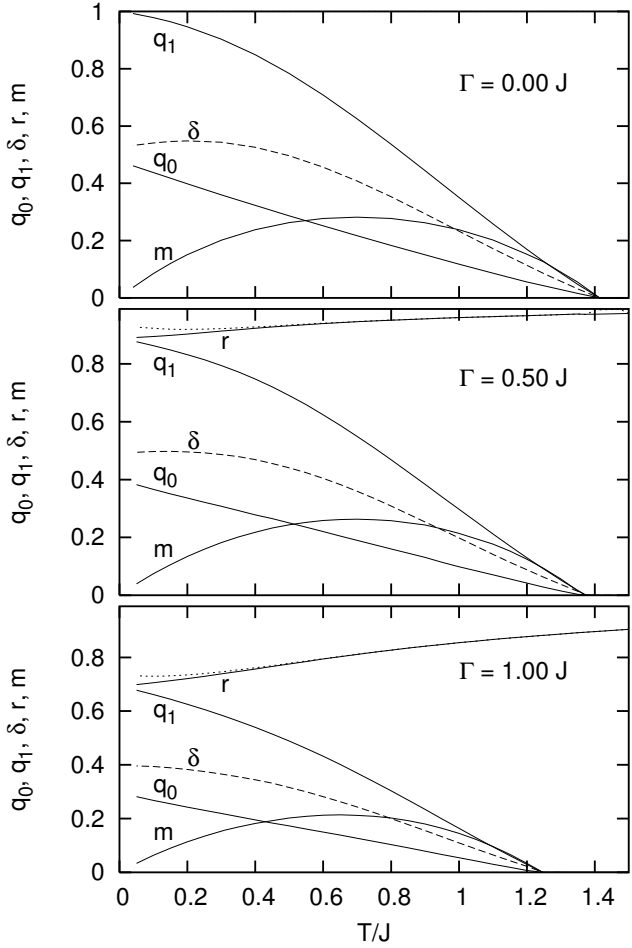}
\caption{Parameters $q_0$, $q_1$, $\delta=q_1-q_0$, $r$ and $m$ as a function of $T/J$ 
for the 2S model. The upper, the middle and the lower panels are, respectively, 
for $\Gamma/J= 0.0,~ 0.5$ and $1.0$. The dotted lines show the RS solution for $r$.}
\label{f2}
\end{figure}
The 2S model recovers the classical results for 1S-RSB \cite{parisi} when $\Gamma=0$.
When $\Gamma>0$, the replica-diagonal correlation $r$ becomes temperature dependent and $r$ must 
be evaluated thogether with the other parameters ($q_0,~ q_1,~ m$).
The results are shown in Fig. \ref{f2}. 
The parameters  $q_0,~q_1$, $m$ and $\delta=q_1-q_0$ (which gives 1S-RSB parameter) 
are zero at $T>T_c$ where the replica symmetry solution is valid.
The transverse field reduces $T_c$ and 
the order parameters  as well. 
The thermodynamic quantities are less sensitive on variations of $m$ at
$T$ near $T_c$ than at lower temperatures. 
For example,
the results obtained using  the
original Parisi's scheme show that $\delta$ increases when $T$ decreases from $T_{c}$ in 
agreement with 
Ref. \cite{eduardo}.  This occurs because right below the transition temperature 
the remaining RSB order parameters have very weak dependence on the solution of 
parameter $m$. 
On the other hand, at lower temperature, the original Parasi's scheme shows 
that the parameter $m$ becomes increasingly important. For instance,
the parameter $\delta$  is
greater than zero 
and the difference between $r$ and $r_{RS}$
(for $\Gamma > 0$) is increased 
when $T$  decreases
by contrast with
Ref. \cite{eduardo} where $\delta$ becomes zero when $T\rightarrow 0$.
Therefore, our results indicate that the validity range in temperature of RSB procedure 
proposed in Ref. \cite{eduardo} is restrained to a region quite close  
to the critical temperature $T_{c}$.  
As consequence, any conclusion 
about whether or not the RS is restored 
in Ref. \cite{eduardo} is
limited by the procedure itself. 
On the contrary, our results obtained by the original Parisi's scheme (in the same static approximation) suggest  
that the RSB effects are present even at $T\rightarrow0$ as it can be seen by the 
extrapolation of curves in Fig. 1 to $T\rightarrow0$.  
\begin{figure}[t!] 
\includegraphics[angle=270, width=8.5cm]{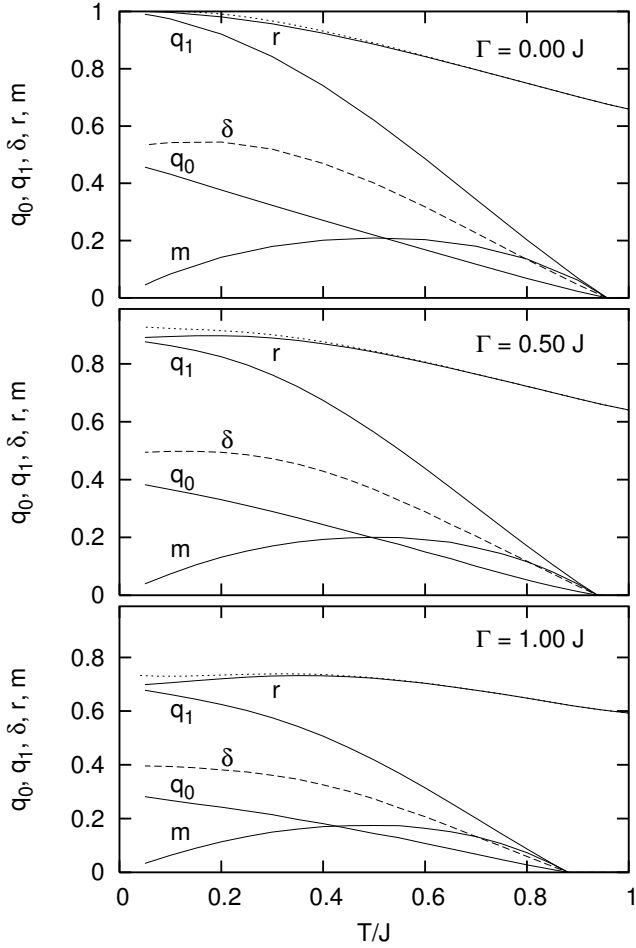}
\caption{Parameters $q_0$, $q_1$, $\delta=q_1-q_0$, $r$ and $m$ as a function of $T/J$ 
for the 4S model. It is used the same convention as Fig. 1 for the dotted lines and values of $\Gamma$.}
\label{f4}
\end{figure}

Fig. \ref{f4} shows the order parameters for the 4S model. 
The behavior of $q_0,~q_1$, $\delta$ and $m$ is qualitatively the same as that one of the 2S model. 
However, $r$ is temperature dependent for all values of $\Gamma$ due to the presence of non-magnetic states. 
On the other hand, the ground state energy of 
the 4S model when $\Gamma=0$ is the same as that of the classical 2S model \cite{sherrington}. 
The results of the 2S and the 4S models are converging to the 
same values when $T\rightarrow 0$ even for $\Gamma>0$. Therefore, it 
could be expected that in the ground state of the 4S model every site carry a 
single fermion. Hence the empty and the double occupied sites do not contribute to the thermodynamic quantities at $T\rightarrow0$.

\begin{figure}[t!] 
\includegraphics[angle=270, width=8.5cm]{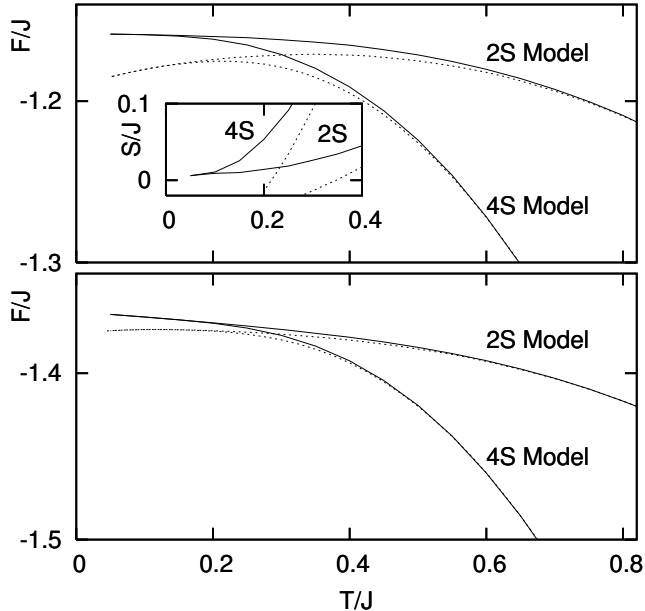}
\caption{Free energy as a function of $T/J$ for the 2S and 4S models. The upper and the 
lower panels are, respectively, for $\Gamma/J= 0.5$ and $1.0$.
The solid lines are results for the 1S-RSB, while the dotted lines are for the RS approach.
The insert shows the entropy at lower temperatures.}
\label{figfree}
\end{figure}
The free energies for 2S and 4S models when $\Gamma=0.5J$ and $1.0J$ are exhibited in 
Fig. \ref{figfree}. The 1S-RSB can be compared with the RS solution. The difference between 
these two approaches increases when $T$ decreases from $T_c$. The 1S-RSB ansatz gives 
higher values for the free energy, and its entropy ($S=-\partial F/\partial T$) is positive 
for the 
range of  temperatures analyzed in Fig. \ref{figfree}. The increase of $\Gamma$ reduces the difference
between the RS and the 1S-RSB.
It could 
be conjectured that when the strength of $\Gamma$ is enhanced quantum fluctuations assume a relevant role
inducing tunneling between the metastable spin glass 
valleys.
Thereby, it could reduce the importance of the replica 
symmetry breaking.
Nevertheless, it is still not able to restore the RS solution. 
Although these results are limited by the static ansatz that is 
expected to yield inaccurate quantitative results at very low temperature, they are 
in good qualitative agreement with those obtained by the Trotter-Suzuki formalism 
without the static ansatz \cite{goldschmidt,buttner1} when $T$ decreases from $T_c$ and by the Landau theory of quantum spin glass rotors \cite{sachdev} when $\Gamma$ increases 
towards the QCP.

To conclude, it is studied the fermionic Ising spin glass model in a transverse field $\Gamma$
within a Grassmann path integral formalism.
The static ansatz and the replica trick with 
one-step replica symmetry breaking scheme are used.
Recently, this problem has been analyzed by the same formalism and the same static approximations, but within a theory
in which the Parisi block size parameter $0<m< 1$ is taken as an independent
parameter of temperature \cite{eduardo}. 
This method in Ref. \cite{eduardo} for finite and fixed values of $m$ 
leads to 
the conclusion that the
RSB parameter decreases towards zero at low temperature and the RS is restored at $T=0$ when
$\Gamma>0$. 
We have compared the procedure proposed in Ref. \cite{eduardo} with the original 
Parisi's scheme which indicates
that this alternative procedure is reliable mainly in the vicinity of the 
critical temperature. Therefore, any extrapolation to lower temperature is limited 
by the procedure itself besides the static approximation.
Moreover, when quantum fluctuations become important, our results provide evidences that the original Parisi's RSB scheme 
should be
used to improve the physical 
description of the ordered spin glass phase treated by the replica method.

This work was partially supported by the Brazilian agency CNPq. 

\end{document}